\documentclass[3p]{elsarticle}

\usepackage{graphicx}
\usepackage{amssymb}
\usepackage{amsmath}
\usepackage[section]{placeins}
\usepackage{etoolbox,lineno}
\usepackage{textcomp}
\usepackage{hyperref}
\usepackage{url}

\hypersetup{colorlinks=true,linkcolor=blue, linktocpage}

\begin{document}

\let\today\relax

\raggedbottom

\begin{frontmatter}

\journal{Physica A: Statistical Mechanics and its Applications}

\title{Predictive intraday correlations in stable and volatile market environments: Evidence from deep learning}

\author[first]{Ben Moews\corref{corresponding}}
\ead{b.moews@ed.ac.uk}
\cortext[corresponding]{Corresponding author}

\author[second]{Gbenga Ibikunle}
\ead{gbenga.ibikunle@ed.ac.uk}

\address[first]{Institute for Astronomy, University of Edinburgh, Royal Observatory, Edinburgh, EH9 3HJ, UK}
\address[second]{Business School, University of Edinburgh, 29 Buccleuch Pl, Edinburgh, EH8 9JS, UK}

\date{Declarations of interest: none}

\begin{abstract}
Standard methods and theories in finance can be ill-equipped to capture highly non-linear interactions in financial prediction problems based on large-scale datasets, with deep learning offering a way to gain insights into correlations in markets as complex systems. In this paper, we apply deep learning to econometrically constructed gradients to learn and exploit lagged correlations among S\&P 500 stocks to compare model behaviour in stable and volatile market environments, and under the exclusion of target stock information for predictions. In order to measure the effect of time horizons, we predict intraday and daily stock price movements in varying interval lengths and gauge the complexity of the problem at hand with a modification of our model architecture. Our findings show that accuracies, while remaining significant and demonstrating the exploitability of lagged correlations in stock markets, decrease with shorter prediction horizons. We discuss implications for modern finance theory and our work's applicability as an investigative tool for portfolio managers. Lastly, we show that our model's performance is consistent in volatile markets by exposing it to the environment of the recent financial crisis of 2007/2008.
\end{abstract}


\end{frontmatter}

\nolinenumbers

% Main body start
\section{Introduction}
\label{sec:introduction}

Changes in stock markets are fuelled by human decisions based on beliefs about a stock's future performance. Investors anticipating the actions of other investors, especially in a period of high uncertainty, also influence such beliefs \cite{Drakos2004, Perryman2010}. This makes markets inherently noisy and prone to fluctuations. Such inconsistencies in the valuation of stock prices has been the subject of a long-standing academic debate centred on the efficient market \cite[see][]{Fama1965, Fama1970} and random walk hypotheses \cite[see][]{Kendall1953, Cootner1964, Malkiel1973}, meaning whether such lagged correlations in the stock markets exist at all. Ferreira and Dion\'iso \cite{Ferreira2016} deliver evidence against the efficient market hypothesis in the U.S. stock market, identifying market memory in the form of correlations over seven months, with related research finding evidence for long-term correlation in market indices \cite{Li2015}.

Previous research establishes the existence of lagged correlations in non-volatile market environments for day-to-day forecasts when combined with infinite impulse response filtering in the data preprocessing. These inputs can be used to realise above-average accuracies in predicting price trend changes without the inclusion of data from the target stock as an input, delivering evidence against the random walk hypothesis and most forms of the efficient market hypothesis in stable market environments \cite{Moews2019}. The growing interest in research dealing with the use of artificial neural networks for stock market prediction is facilitated by the availability of large historical stock market trading data \cite{Huck2009, Chong2017, Heaton2017, Krauss2017}. Such data usually takes the form of time series, and thus classical approaches to time series analysis are currently widespread within the investment industry \cite{Clarke2000}. This configuration, together with the existence of related hypotheses, makes the prediction of price changes based on historical data an attractive use case for trend forecasting involving inter-correlated time series in stock markets as an example of real-world complex systems \cite{Johnson2003}.

We investigate intraday predictability with differing time intervals, and test our model's complexity and performance in high-volatility crisis scenarios. We find evidence of the presence of time-delayed correlations in S\&P 500 stocks in both stable and volatile markets, and of the viability of using deep learning for trend predictions in large numbers of inter-correlated time series. Our experiments outperform predefined baselines for strict statistical key performance indices, which includes accuracies for different prediction horizons. Predictions of one stock's trend changes based on other stocks' price trend gradients in the preceding time step show an improved accuracy for larger time intervals, with average and maximum accuracies of 56.02\% and 63.95\%, respectively, for one-day predictions. In this regard, we outperform related research by Huck \cite{Huck2009} and Krauss et al. \cite{Krauss2017}, and slightly underperform for day-to-day predictions when compared to the results by Moews et al. \cite{Moews2019}, demonstrating the negative impact of a lack of input smoothing. 

Our findings are novel in that we exploit correlations of a target stock with other stocks in the same economy under different market scenarios, without recourse to industry classifications and input smoothing, to make successful predictions about the target stock's price evolution. Furthermore, we address intraday intervals, which are more relevant in today's high-tech high frequency trading market environments. The results are consistent in financial crisis situations and demonstrate that our framework is able to exploit existing and inter-temporal correlations in a highly volatile market environment. In addition, we gauge the complexity of the targeted prediction problem by forcing information compression through bottleneck layers of different sizes. With this work, we further demonstrate the utility of linear regression derivatives as inputs for the investigation of economic hypotheses and risk diversification with deep learning.

Our results deliver rigorously tested empirical evidence against the random walk hypothesis, as the assumption of stock prices over time as random walks effectively excludes the possibility of exploitable information in historical stock market data. The findings thus contradict Sitte and Sitte \cite{Sitte2002}, who argue in favour of the existence of a random walk due to the perceived inability of artificial neural networks to extract actionable information. Our results agree with previous results for one-day predictability in stable market environments, adding to the evidence against most forms of the efficient market hypothesis \cite{Skabar2013, Moews2019}. Previous research also shows that market efficiency is highly dependent on how developed a stock market is, with higher-developed markets exhibiting increased efficiency \cite{Rizvi2014}.

The remainder of the paper is set out as follows: Section~\ref{sec:prediction} discusses the progress made on predicting stock returns and prices, and the functionality of deep learning models in finance. Section~\ref{sec:data_and_methodology} presents the data and methodology used in this paper, Section~\ref{sec:results} discusses the results and findings of our experiments, and Section~\ref{sec:conclusion} concludes our work.

\section{Predicting stock returns}
\label{sec:prediction}

\begin{figure}[!htb]
\begin{center}
% NOTE: FIGURES ARE OPTIMISED TO NOT SPAN THE FULL TEXT WIDTH (1.5-COLUMN FIT)
% NOTE: FIGURE SHOULD NOT BE PRINTED IN COLOUR
\includegraphics[width=0.7\columnwidth]{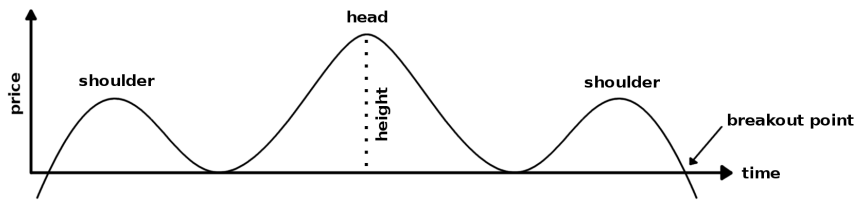}
\caption{Head-and-shoulders pattern in stock market data is not a mathematically developed decision aid. However, the pattern is widely used by technical analysts as a guideline for investments, as it is perceived to follow a predictable scheme as a rule-of-thumb within the stock market.}
\label{fig:figure_1}
\end{center}
\end{figure}

Technical analysis involves making stock trading decisions on the basis of historical stock market data. The assumption behind its utilisation in the investment industry is that above-average risk-adjusted returns are possible when using past time series of stock information, and Clarke et al. \cite{Clarke2000} show that this practice is wide-spread in the investment industry. A meta-analysis by \cite{Park2004} also demonstrates that the majority of studies on the topic of technical analysis report a profitability that undercuts the arguments of the efficient market hypothesis \cite{Tanaka-Yamawaki2007}. Apart from moving average methods, perceived patterns are among the other, less investigated methods that are used by technical analysts, for example the so-called ``head-and-shoulders pattern''. The latter is used as an indicator for a trend shift, using the negative value of the height denoted in Figure~\ref{fig:figure_1} as the target price for a trade initiated at the breakout point, which marks the pattern's completion.

The lack of statistical research on such patterns has been criticised by \cite{Neftci1991}, noting that there is a disparity between the rigour of academic time series analysis and the decision-making of traders. Subsequent studies by \cite{Chang1999} and \cite{Lo2000} show indicators of applications for select currencies in foreign exchange markets and conclude that such patterns may hold some practical value. The above explanations are, however, only provided for the sake of a holistic overview, and this paper does not delve further into, or apply, technical analysis.

White \cite{White1988} hypothesised early that artificial neural networks could be successfully used to deliver empirical evidence against all the three forms of the efficient market hypothesis, reporting an $R^2$ value of 0.175 when using a simple feed-forward network with five previous days of IBM stock prices as inputs for a regression task \cite[see also][]{Saad1998}. More recently, Dixon et al. \cite{Dixon2017} also implement an artificial neural network with five hidden layers for trinary classification, differing in an output that represents little or no change from the previously cited studies. Using data of CME-listed commodities and foreign exchange futures in five-minute intervals to generate a variety of engineered features like moving correlations, a single model is trained instead of a separate model for each target instrument, resulting in an average accuracy of 42.0\% for the investigated three-class prediction task.

No cross-validation is carried out in the above work, which would further validate the results for economic conclusions. The latter process refers to the splitting of the dataset into multiple sets of approximately the same size. A model is then trained on all but one of the sets and tested on the remaining one, alternating which sets are used for training and testing. This way, multiple tests are carried out over all instances of the data, and the results for all models are then averaged. This is the ``gold standard'' in machine learning, as it addresses the possibility of testing models on an unrepresentative subset. Cross-validation makes for more reliable and stable results; therefore, we apply this approach to our analysis \cite{Arlot2010}.

Krauss et al. \cite{Krauss2017} investigate the effectiveness of deep learning, gradient-boosted-trees and random forests in the detection of statistical arbitrage. Their results indicate that, when combining one each of deep-layer neural networks, gradient-boosted trees and random forests, one may obtain returns exceeding 0.45\% per day before taking transaction costs into account. Research on artificial neural networks for stock market prediction does, however, remain sparse over the last two decades. In contrast, momentum trading strategies have received increased attention in financial research in recent times. The apparent ability of momentum-based strategies to outperform the market are viewed as a premier anomaly within the framework of the efficient market hypothesis \cite[see][]{Fama2008}.

Another area of research that has gathered attention is the text-based prediction of stock markets using machine learning models. The notion of using news articles, with new information, as opposed to historical market data, to predict stock prices was introduced by Lavrenko et al. \cite{Lavrenko2000} and is a common baseline for subsequent research. A system devised by Schumaker and Chen \cite{Schumaker2009a}, named AZFinText, which employs wide-spread news coverage, results in a directional accuracy of 57.1\% for the best-performing model. The approach involves a support vector machine with a proper-nouns scheme instead of a simple bag-of-words approach, in combination with a stock's current price as inputs, over a five-week period. The authors of the above research show their system to be successful within a twenty-minute time frame, which falls under the margin of earlier research concluding that the majority of market responses to new information experiences a time lag of approximately ten minutes \cite[see][]{Patell1984}. Subsequent research demonstrates that AzFinText can outperform established quantitative funds \cite{Schumaker2009b}. In related research, Ding et al. \cite{Ding2015} propose the use of a neural tensor network to learn event embedding from financial news articles in order to feed that information into a deep convolutional neural network for a two-class prediction of a stock price's future movement. For this system, an accuracy of 65.9\% is reported for 15 different S\&P 500 stocks and daily trend predictions. No clear indication, however, is given as to how the reported stocks are selected, which undermines the reported results.

\section{Data and methodology}
\label{sec:data_and_methodology}

\subsection{Obtaining suitable datasets}
\label{datasets}

We obtain two sets of high-frequency transaction data from the Thomson Reuters Tick History \footnote{\url{https://developers.refinitiv.com/thomson-reuters-tick-history-trth}} database. Each dataset contains a stock identification code, the transaction date and time to the nearest millisecond, the high/best bid price, the low/best ask price, the volume, and the average transaction price. The first dataset is sampled from April 2011 to April 2016, covering five years of transaction price information for S\&P 500 stocks. The second dataset spans April 1996 to April 2016, which results in a larger dataset covering 20 years, including the financial crisis period of 2007/2008. The datasets show instances of missing values for the price variables in some stocks, especially earlier in the largest of the datasets, thus suggesting that there are no transactions recorded for those intervals. We follow Chordia et al. \cite{Chordia2001} by replacing the missing values with the same-column entries of the preceding index or the index with the next non-missing value, depending on whether the former belongs to the same stock as indicated by the stock identification code. In order to generate feature vectors that can be used as inputs, the time stamps have to be aligned perfectly. We implement a data cleansing approach to secure a time-wise alignment of observations for different stocks by substituting missing rows. 

As na\"ive procedures that loop over complete datasets and copies of a full matrix for every missing observation result in infeasible time estimates, our approach acts solely on time vector comparisons for a specific stock's matrix, with matrix shifts in the case of local time vector incompatibility, and operating on a declared matrix of sufficient dimensionality to allow for insertions instead of appending values. All observations with inexplicable and incomplete timestamps are also eliminated. Following the data cleansing process, the datasets are checked for a subset of stocks satisfying the requirement of being consistently present over a sufficiently large portion of the dataset's time frame divisible by the chosen number of time steps for the subsequent gradient calculation. The datasets' price information is extracted and transformed into a feature matrix with row-wise time steps and column-wise stocks.

\subsection{Statistical feature engineering}
\label{sec:feature_engineering}

Feature engineering describes the manual selection and, if necessary, transformation of given datasets into new data that better represents the features needed for a chosen task. For this paper, and following Moews et al. \cite{Moews2019}, simple linear regressions are used as way to approximate the trends over given time intervals. These regressions are least-squares estimators of such models with one explanatory variable to fit a line that minimises the squared sum of the residuals. They take the form of a minimisation problem to find the intercept $\beta_0$ and the slope $\beta_1$,
\begin{equation}
\underset{\beta_0, \beta_1}{\min} \ Q(\beta_0, \beta_1) = \underset{\beta_0, \beta_1}{\min} \left( \sum_{i=1}^n (y_i - \beta_0 - \beta_1 x_i)^2 \right).
\smallskip
\end{equation}
By running a linear regression over each time series and time interval separately, and by taking the first derivative of the resulting equation, the trend gradients for single stocks and time intervals are obtained. Given aligned stock prices for $N$ points in time and a chosen step size $s$, the resulting feature vector generated from a stock's price has the length $N/s$. Depending on the time frame covered by the dataset, limits apply to the size of intervals that can be chosen to still obtain a viable size for the training set. For the five-year dataset, three sets of gradients are computed: The first set covers 1,242 one-day gradients per stock, the second set covers 7,361 one-hour gradients, and the third set covers 14,725 half-hour gradients for 449 stocks. For the 20-year dataset with hourly values, daily gradients are computed for the years from 2003, preceding the global financial crisis, to and including 2008, resulting in 2,131 gradients.

\subsection{Training deep learning models}
\label{sec:training}

\begin{figure}[!htb]
\begin{center}
% NOTE: FIGURES ARE OPTIMISED TO NOT SPAN THE FULL PAGE WIDTH (1.5-COLUMN FIT)
% NOTE: FIGURE SHOULD NOT BE PRINTED IN COLOUR
\includegraphics[width=0.7\columnwidth]{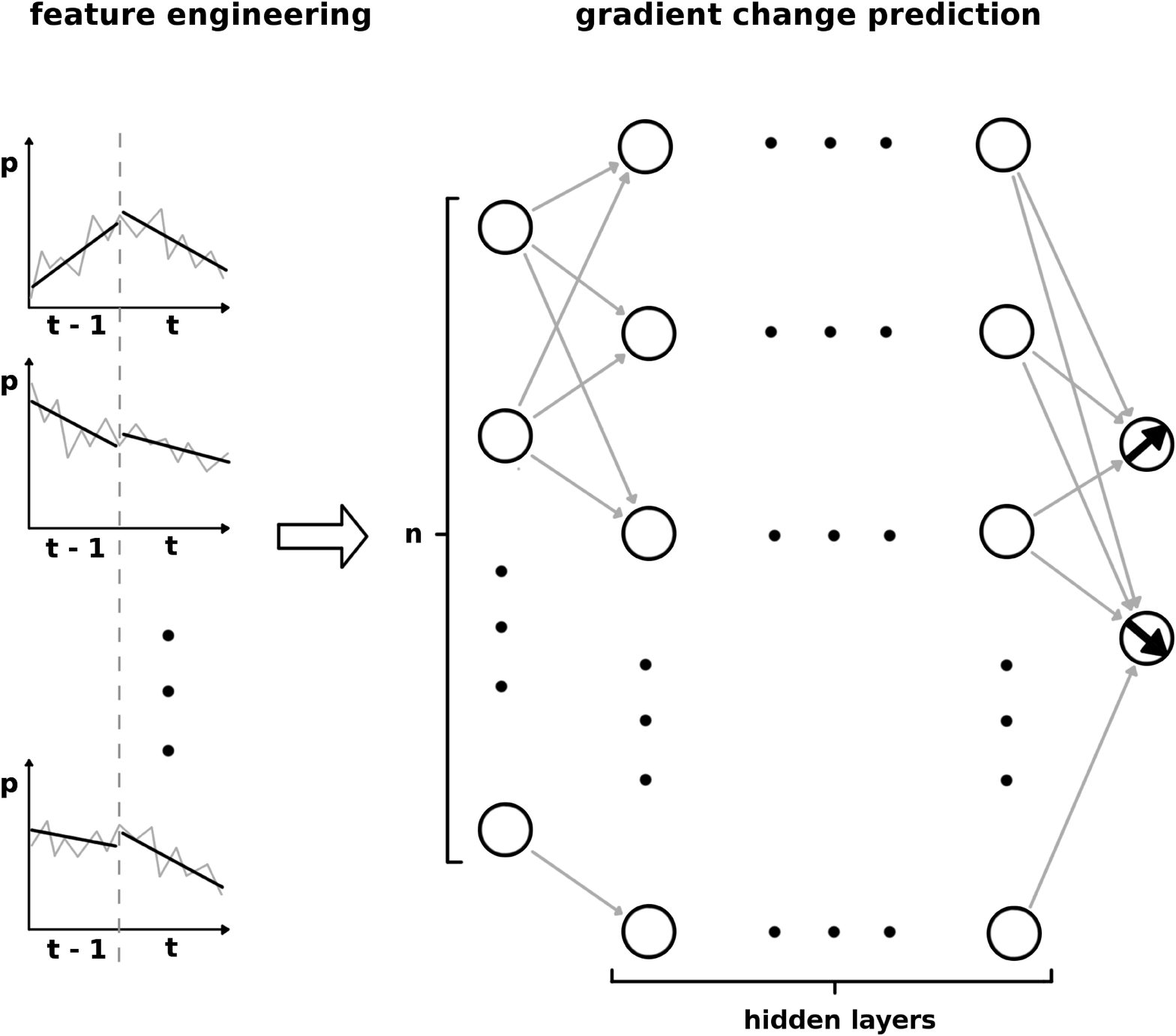}
\caption{Model setup for the experiments. A simple linear regression is performed on each time step for each stock's price series, after which the first derivative of the resulting regression is computed as a trend strength indicator. For each target stock, all stocks' gradients for the preceding time step except for information about the target stock are then used as input features for an neural network with five hidden layers and two output nodes for a binary classification task. The model is then, with separate experiments for each stock, trained to predict the upwards or downwards change of the target stock's trend gradient in the next time step.}
\label{fig:figure_2}
\end{center}
\end{figure}

Figure~\ref{fig:figure_2} depicts a schematic overview of the experimental setup used in this paper. For the number $n1$ of stocks that are made usable during the data cleansing and preprocessing, gradients of the price trends for each separate stock are computed in the feature engineering step, as described in Section~\ref{sec:feature_engineering}. The $n - 1$ gradients for one time step $t - 1$ are then used as inputs to a feedforward artificial neural network that is fully connected for adjacent layers to predict whether the gradient of the left-out $n^\mathrm{th}$ stock changes upwards or downwards with regard to its gradient in the preceding time step $t - 1$. This setup resembles Moews et al. \cite{Moews2019} and allows the comparison of one-day results with the latter. The reason for omitting the target stock gradient for the preceding period in the input vector is that the described framework is designed to test for lagged correlations, which prohibits using the target stock's own information. Predictions are meant to explore the predictability based solely on correlations with other members of the S\&P 500 stocks. This is a major distinguishing aspect of this paper from related research on time series-based stock market prediction.

In order to identify general correlations, we reduce result variability via five-fold cross-validation  After each five-way split and sorting into a training set of 80\% of the dataset for the respective fold, another 25\% of the training set is partitioned off as the validation set for an early stopping procedure, resulting in a 60-20-20 split for each fold and stock. The experiments are run for all stocks, with an output vector that is replaced by a binary one-hot representation to state whether a gradient in a given time interval is increased or reduced for the next interval. Normalising the inputs is necessary in order to address geometrical biases, to distribute the importance of values equally, and to ensure that all values are situated in the same range to make them comparable for an efficient learning process. The training examples split from the dataset are normalised element-wise via
\begin{equation}
\mathbf{X}_{norm} = \frac{\mathbf{X} - \mathbf{X}_{min}}{\mathbf{X}_{max} - \mathbf{X}_{min}}.
\end{equation}
Regularisation is a valuable way to address issues with overfitting, meaning poor generalisation. Commenting on the inherent vagueness of parameter tuning with regard to real-world applications, Snoek et al. \cite{Snoek2012} summarise the problems associated with the terrain by stating that machine learning algorithms ``[...] frequently require careful tuning of model hyperparameters, regularization terms, and optimization parameters. Unfortunately, this tuning is often a `black art' that requires expert experience, unwritten rules of thumb, or sometimes brute-force search'' \cite{Snoek2012}. While such a statement sounds bleak, the parameters and hyperparameters that have to be set can be determined, or at least approximated. The choices we make in this study follow a combination of established scientific reasoning and preliminary experimentation, as well as experience in the application of deep learning architectures and simple heuristics. These are employed to slowly increase the model's complexity, and further details can be found in Nielsen \cite{Nielsen2015}. Preliminary tests on a subset of stocks result in a model architecture with five hidden layers, with 400 artificial neurons each.

Addressing potential memory issues, a mini-batch size of 100 is chosen, and each model is trained for 50 epochs, with early stopping as a regularisation measure. In order to optimise the gradient descent trajectory, our setup also utilises dynamic learning rate decay and momentum. We make use of hyperbolic tangent activations, sigmoid outputs, and Gaussian-initialised model weights \cite[see][]{He2015}. In training artificial neural networks, sigmoid functions are a term applied to the special case of the logistic function, with a steepness value of $k = 1$ and a midpoint of $x_0 = 1$. The sigmoid function is calculated as
\begin{equation}
\mathrm{sigm}(\mathbf{x})_j = \frac{1}{1 + e^{-k \cdot (x_j - x_0)}} .
\end{equation}

It is important to note that values for the sigmoid function level out at zero on the lower end, which can lead to a fast saturation of weights at the top layers of multi-layered artificial neural networks \cite{Glorot2010}. For this reason, we restrict its application to the output layer and use the hyperbolic tangent function for activations, which is similar but is centred on zero instead of 0.5, with a lower limit of $-1$ and the same upper limit of one, meaning
\begin{equation}
\begin{split}
\mathrm{tanh}(\mathbf{x})_j = \frac{\mathrm{sinh}(x_j)}{\mathrm{cosh}(x_j)} = \frac{e^{x_j} - e^{-x_j}}{e^{x_j} + e^{-x_j}} = \frac{1 - e^{-2x_j}}{1 + e^{-2x_j}} .
\end{split}
\end{equation}

In addition, $\ell_2$ regularisation is chosen over $\ell_1$ regularisation. This decision is driven by the encouragement to use all inputs to a certain degree, as a complex interdependence of the studied time series is assumed due to the failures of past approaches to identify simpler correlations. Through the introduction of decaying weights, and with the parameter $\lambda$ defining the trade-off between the loss function and the penalty for larger weights, the gradient descent is given as
\begin{equation}
w_{j, i} \ = \ w_{j, i} - \eta \ \frac{\partial E}{\partial w_{j, i}} - \eta \lambda w_{j, i},
\end{equation}
where $w_{j, i}$ is a weight, $\eta$ is the chosen learning rate, and $E$ is the total error valued on the basis of loss functions under a certain set of weights between layers. Using the example of the quadratic cost function, the total error for this case can be calculated as $E = 0.5 \sum_i \sum_j (\hat{y}_{j, i} - y_{j, i})$, where $j$ indexes the output units and $i$ the pairs of training examples and corresponding outputs, and $\hat{y}$ and $y$ denote the calculated outputs and actual labels, respectively.

\subsection{High volatility and bottlenecks}
\label{sec:high_volatility}

In the context of the given problem, an interesting question is that of its general complexity with regard to the number of variables the relevant data can be reduced to for an acceptable accuracy. We investigate this by inserting a bottleneck layer consisting of a small number of neurons into the models for the daily predictions based on the five-year dataset with hourly values. This process is implemented for numbers of bottleneck nodes from the set $\{1, 3, 5, 10 \}$ to see how the bottleneck size influences the accuracy. This approach is favoured over using autoencoders \cite[see][]{Heaton2017} as a precedent to using their compression layer as inputs for a full model, as autoencoders learn a goalless reduction of their inputs.

\begin{figure}[!htb]
\begin{center}
% NOTE: FIGURES ARE OPTIMISED TO NOT SPAN THE FULL PAGE WIDTH (1.5-COLUMN FIT)
% NOTE: FIGURE SHOULD NOT BE PRINTED IN COLOUR
\includegraphics[width=0.7\columnwidth]{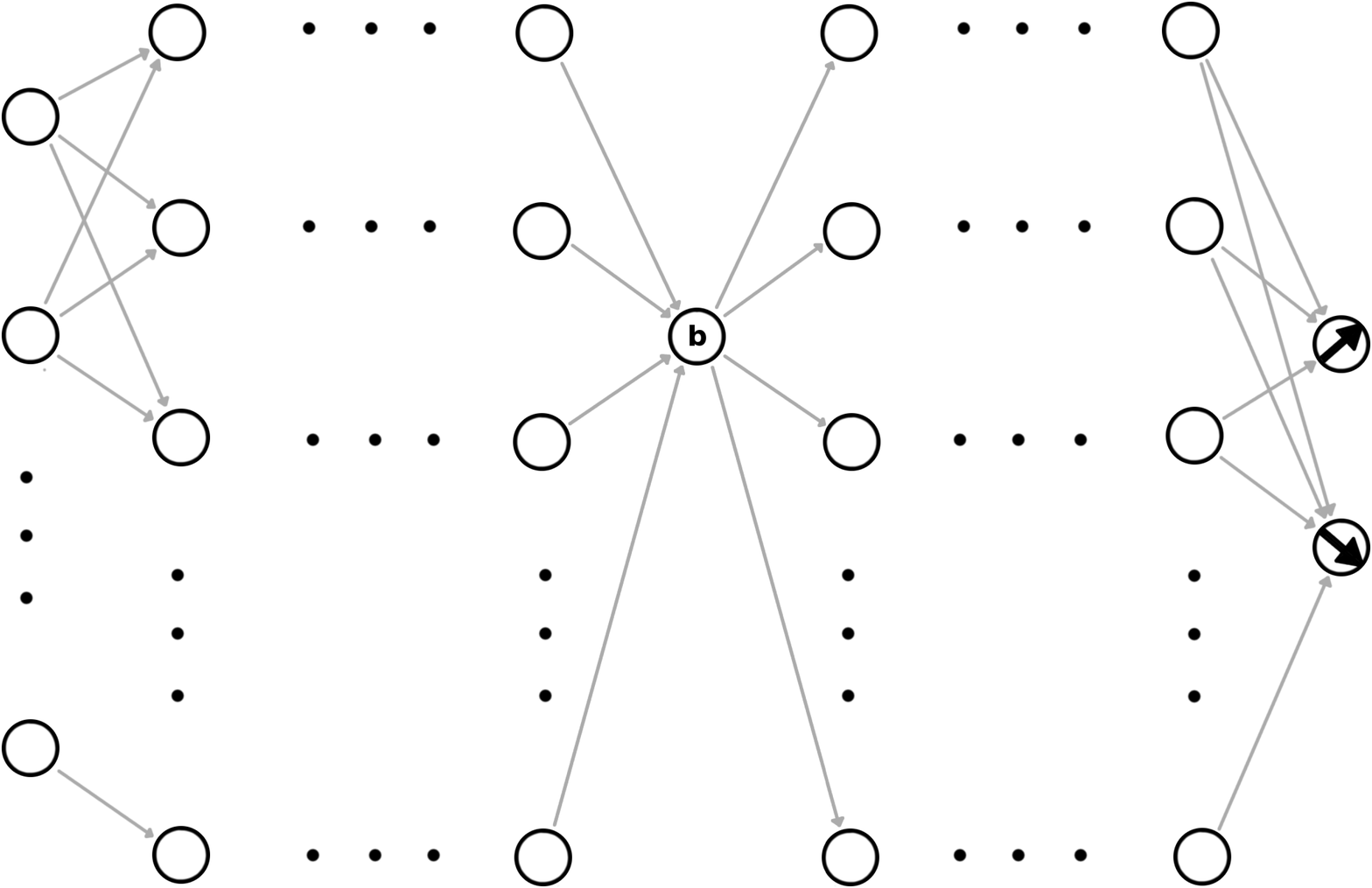}
\caption{Model with a one-neuron bottleneck layer. The depicted model inserts a one-node bottleneck in a deep feed-forward neural network model for binary classification. The reason for this insertion is to force the model to learn a goal-oriented reduction of the model's information by ``squeezing'' the necessary information through a narrow representation mid-way.}
\label{fig:figure_3}
\end{center}
\end{figure}

While similar to using a bottleneck, auto-encoders recreate their input as their output through a reduction layer without a target outside of the input recreation, whereas our chosen method reduces the complexity via the bottleneck layer with the explicit goal of making that reduction suitable for the computational task. For a bottleneck layer in the same model, the latter is forced to learn a compressed representation that is suited to the target predictions at hand, funnelling the model's learning process through the nodes of the respective bottleneck. Figure~\ref{fig:figure_3} shows the resulting model, featuring an exemplary additional bottleneck layer for complexity tests with one node marked $\mathbf{b}$.

Virtually all previous studies applying machine learning approaches to stock market prediction, with the exception of Krauss et al. \cite{Krauss2017}, are not tested for above-average results in volatile market situations. However, Schumaker and Chen \cite{Schumaker2009a}, in their development of the AZFinText system, note the importance of such a test. For time series-based prediction, high-volatility environments are more problematic contexts than for text analysis-based systems like AZFinText. This is because they rely solely on stock market information that is effectively in a state of turmoil. In order to test the gradient-based approach and the model implementation for such environments, data from 2003 to, and including, 2008 is extracted from the 20-year dataset with hourly values and used to compute daily gradients. No cross-validation is performed for these experiments, as the test set has to represent a phase of enhanced volatility in the market.

In order to achieve this, a test set from July 2007 to the end of 2008 is split from the set of gradients, with the previous 4.5 years serving as training data. This setup also more closely resembles an applicable prediction system instead of just an economic hypothesis testing framework, as only past data is used instead of identifying general correlations across different time periods through cross-validation. Due to the large fluctuations in the dataset, and given that some stocks remain more stable than others during the financial crisis of 2007/2008, a higher variance of accuracies is expected. An accuracy above the baseline, which is expected to be higher than random chance due to the general negative trend in that time period, delivers evidence for the persistence of correlations in high-volatility scenarios such as global crises.

\subsection{Reliability and robustness analysis}
\label{reliability}

In order to validate our results, baselines addressing the market behaviour and the distributions of target vectors are necessary to find statistically significant evidence. The focus of this paper on lagged correlations between stocks in relation to economic theory, manifested in excluding the target stock data in the models' inputs, sets this work apart from research that aims to find the best-possible stock market predictions instead of correlations in stock market prices. For research focussed purely on prediction accuracy, baselines representing na\"ive regression (or classification) or basic machine learning methods are more suitable. For related research in Moews et al. \cite{Moews2019}, however, the gradient-based deep learning approach is shown to considerably outperform a simple support vector machine, which translates to our work. In this paper, the threat of coincidences is partly approached through cross-validation, but a model's accuracy must also lie significantly above the accuracies of one-value and random predictions.

Another area of concern relates to the accuracy of random mock predictions, which we address by randomly shuffling the predictions of our model. If the accuracies of these mock predictions are similar to the accuracies achieved through unshuffled predictions, this indicates that only distributional characteristics are learned. Importantly, this shuffled set of mock predictions also provides a test against the alternative explanation of market shocks, as correlations resulting from the latter would move more uniformly and, as such, would lift this comparison baseline closer to the model accuracy. The third problem that needs to be contended with is that models could identify the majority class of the binary targets, simply outputting the latter. This potential issue can be explored by creating two additional sets of mock predictions, each one covering one of the binary targets, which also allows a comparison regardless of target imbalances. The model's predictions have to outperform both of these mock sets in order to demonstrate the results not being the result of simply learning the majority class.

The last baseline is a combination of the three described baselines, for which we take the highest accuracy for each stock and time step. In doing so, the resulting set of predictive accuracies is guaranteed to be always be at least 50\% and provides an easy way to check for model performance against a single baseline. We calculate the average accuracy across S\&P 500 stocks for a given experiment, and visually compare it to the baseline results via notched box-and-whiskers plots. In addition, we also provide the lower bounds for confidence intervals and $p$-values for an upper-tail Welch's $t$-test with $a = 0.001$.

\section{Results and discussion}
\label{sec:results}

\subsection{Primary experiments}
\label{sec:primary_results}

\begin{figure}[!htb]
\begin{center}
% NOTE: FIGURE SHOULD SPAN THE FULL PAGE WIDTH (2-COLUMN FIT)
% NOTE: FIGURE SHOULD NOT BE PRINTED IN COLOUR
\includegraphics[width=\columnwidth]{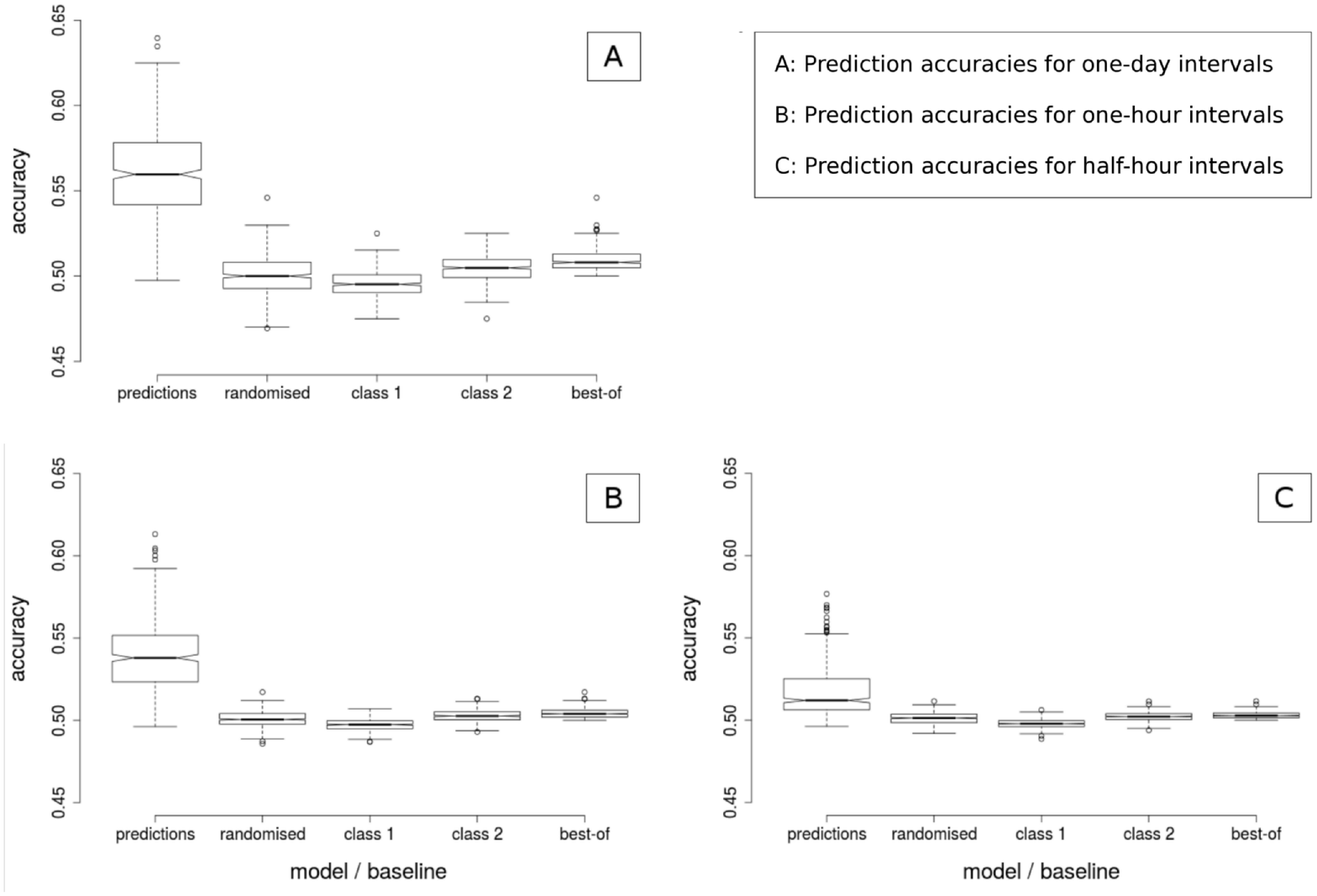}
\caption{Notched box-and-whiskers plots for accuracies of various time intervals. Panels A, B and C present the plots for one-day, one-hour and half-hour intervals, respectively. On the horizontal axis, `class 1' and `class 2' concern downwards-only and upwards-only changes, `randomised' represents randomly shuffled model predictions in order to test for simple distribution learning, and `best-of' takes the best result out of all three baselines at each step. A statistically significant difference in means at 95\% confidence is indicated if notches are non-overlapping. The lower and upper ends of a box indicate the first and third quartile, while the median is depicted as a horizontal bar. Outliers are shown as circles above or below the whiskers.}
\label{fig:figure_4}
\end{center}
\end{figure}

In panels A, B and C of Figure~\ref{fig:figure_4}, we present box-and-whiskers plots to visualise the results of the one-day, one-hour and half-hour gradient interval results, respectively. For every experiment, the accuracies for the model and the baselines are given, as well as the $p$-values with regard to the means and the minimal difference for a 99.9\% confidence interval. The whiskers show the lowest and highest data point that is within 1.5-times the interquartile range of the first and third quartile, and are computed via
\begin{equation}
\begin{split}
\mathrm{whisker}_{\mathrm{\uparrow}} = \min(\max(\mathrm{data}), \ Q_3 + 1.5 \cdot (Q_1 - Q_3)), \\
\mathrm{whisker}_{\mathrm{\downarrow}} = \max(\min(\mathrm{data}), \ Q_1 - 1.5 \cdot (Q_1 - Q_3)).
\end{split}
\end{equation}
The reason for using Welch's $t$-test instead of Student's $t$-test is that the former is more robust when confronted with unequal variances. As shown in Figure~\ref{fig:figure_4}, this choice proves to be relevant due to larger variances of model predictions.
 
The accuracy for one-day predictions, as shown in panel A of Table~\ref{tab:table_1}, is 56.02\% and shows statistically significant deviation in means from the baselines, with associated $p$-values affirming the discardment of the null hypothesis. This result is further reinforced by panel A of Figure~\ref{fig:figure_4} for medians, with no overlap in both notches and boxes in the box-and-whiskers plots. Panel B of Figure~\ref{fig:figure_4} shows the plot for the one-hour gradient intervals. With an accuracy of 53.95\%, the model's accuracies exhibit the same increased variability as for one-day gradients, albeit at a smaller rate. The baselines' accuracies are centred more closely on 50\% (see also panel B of Table~\ref{tab:table_1}), which is consistent with the overall smaller spread of the accuracies for both the model and the baselines, trading accuracy for narrowness. 

Figure~\ref{fig:figure_4}'s panel C, showing the plot for the half-hour interval gradients, suggests a model accuracy of 51.70\%. First, the distribution of the model's accuracies is also skewed towards lower values, meaning more variability above the median. Secondly, there appears to be an emerging pattern when all three plots are considered together, as the model accuracies are lower for smaller gradient time intervals. The narrower boxes for the baselines also imply, however, that the quartile ranges of the accuracy values are narrower, evidencing decreasing variability as considered time intervals become shorter.

\begin{table}[!htb]
\centering
\begin{tabular}{l l l l l l}
\hline
\multicolumn{5}{l}{\textbf{accuracies of predictions}} \\
\hline\noalign{\smallskip}
panel & model & rand. & class 1 & class 2 & best-of \\
\noalign{\smallskip}\hline\noalign{\smallskip}
A & $0.5602$ & $0.5002$ & $0.4955$ & $0.5045$ & $0.5092$ \\
B & $0.5395$ & $0.5008$ & $0.4973$ & $0.5027$ & $0.5043$ \\
C & $0.5170$ & $0.5011$ & $0.4979$ & $0.5021$ & $0.5030$ \\
\noalign{\smallskip}\hline\noalign{\smallskip}
\multicolumn{5}{l}{\textbf{tests against baselines}} \\
\noalign{\smallskip}\hline\noalign{\smallskip}
panel & measure & rand. & class 1 & class 2 & best-of \\
\noalign{\smallskip}\hline\noalign{\smallskip}
A & \textit{p}-value & $< 0.001$ & $< 0.001$ & $< 0.001$ & $< 0.001$ \\
B & \textit{p}-value & $< 0.001$ & $< 0.001$ & $< 0.001$ & $< 0.001$ \\
C & \textit{p}-value & $< 0.001$ & $< 0.001$ & $< 0.001$ & $< 0.001$ \\
A & min. diff. & $0.0599$ & $0.0607$ & $0.0518$ & $0.0471$ \\
B & min. diff. & $0.0356$ & $0.0392$ & $0.0338$ & $0.0322$ \\
C & min. diff. & $0.0137$ & $0.0169$ & $0.0127$ & $0.0118$ \\
\noalign{\smallskip}\hline
\end{tabular}
\caption{449 S\&P 500 stocks are used to predict directional trend changes for each stock, with only the respective other 448 stocks' trends from the preceding time interval as input features. In the table, accuracies, $p$-values and the minimal difference with regard to the significant difference in means are given. Here, `class 1' is the prediction that stock trends will change downwards, and `class 2' is the prediction that stock trends will change upwards, whereas `randomised' represents the same prediction distribution as the model's predictions sampled randomly to test for simple distribution learning. The `best-of' features the respective highest accuracy of the three baselines, meaning `class 1', `class 2', and `randomised', to result in a fourth baseline that is, by default, at least 50\%. Panels A, B and C present the estimates for one day, one hour and half-hour intervals, respectively.}
\label{tab:table_1}
\end{table}

In combination, these metrics deliver strong evidence in support of the economic arguments implied in our hypothesis, meaning that price series in historical stock market data contain lagged correlations that can be successfully exploited with deep-layered feed-forward neural networks, and that the time horizon has a notable effect on predictability. This exploitation can result in above-baseline price trend predictions without using any data from the target stock. Larger time intervals for gradient calculations and predictions based on the latter result in higher average accuracies, but with a trade-off in the form of an increased spread of the accuracies, meaning a larger variance. It therefore seems to be easier for the models to learn correlations between gradients and make corresponding predictions for larger time steps, although more volatility with regard to the variance is incorporated.

A natural explanation for these differences is the presence of more noise in short-time stock observations, indicating that noise is smoothed out for regressions over larger intervals. Such noise events, such as bid-ask bounce, are microstructural in nature and are thus more prevalent in higher-frequency data. Although our results are in direct contradiction to two of the well-established hypotheses in modern finance, the findings presented are not exactly far-fetched. While meta-analyses should always be interpreted cautiously due to the possibility of publication biases, Park and Irwin \cite{Park2004} find that a majority of the published research dealing with quantitative stock market prediction reports findings that indicate a critical view of the efficient market hypothesis in its strong form.

Gradient-based deep learning for stock market trend prediction is explored in Moews et al. \cite{Moews2019}, but employs infinite impulse response filtering to smooth input features and only covers daily predictions in non-volatile markets. Therefore, a further distinction of our study are the periods on which the stock price predictions are based, namely one-day, one-hour and half-an-hour predictions instead of months or single days, as is the case with existing research. Focussing on higher-frequency intervals is critical given the evolution of global financial markets in recent times, as trading now occurs in large volumes at sub-second frequencies and at much lower trade sizes \cite[see][]{Chordia2001}. Thus, investment windows of even significantly large funds have shortened over time. In terms of an investigative tool for portfolio managers, previous studies have explored the applicability of machine learning to intraday risk prediction \cite{Groth2011}. Our approach, in contrast, offers an easily quantifiable way to measure the impact of complex non-linear market correlations of individual stocks. In order to shield portfolios against market movements, the identification of stocks with very low correlations is facilitated via stocks with low to non-existent trend predictability in our framework, thus offering an additional tool for risk diversification.

Our results for daily predictions outperform related research on neural networks for one-day trend predictions \cite{Huck2009, Krauss2017}. In contrast, our result of 56.02\% underperforms in comparison to 58.10\% reported by Moews et al. \cite{Moews2019}, the difference being that no aid in the form of input smoothing is provided to the models. The latter research also answers the question about the impact of target stock information being additionally provided to the model; the result is a minimal increase in accuracy from 58.10\% to 58.54\%, meaning that almost all relevant information is already reflected in, and can be extracted from, other stocks' trend approximations. As our work is, to our knowledge, the first study to compare prediction horizons and model viability in financial crisis environments, we hope that future research will enable comparisons.

\subsection{Complexity and volatile environments}
\label{sec:volatility}

\begin{figure}[!htb]
\begin{center}
% NOTE: FIGURES ARE OPTIMISED TO NOT SPAN THE FULL PAGE WIDTH (1.5-COLUMN FIT)
% NOTE: FIGURE SHOULD NOT BE PRINTED IN COLOUR
\includegraphics[width=0.7\columnwidth]{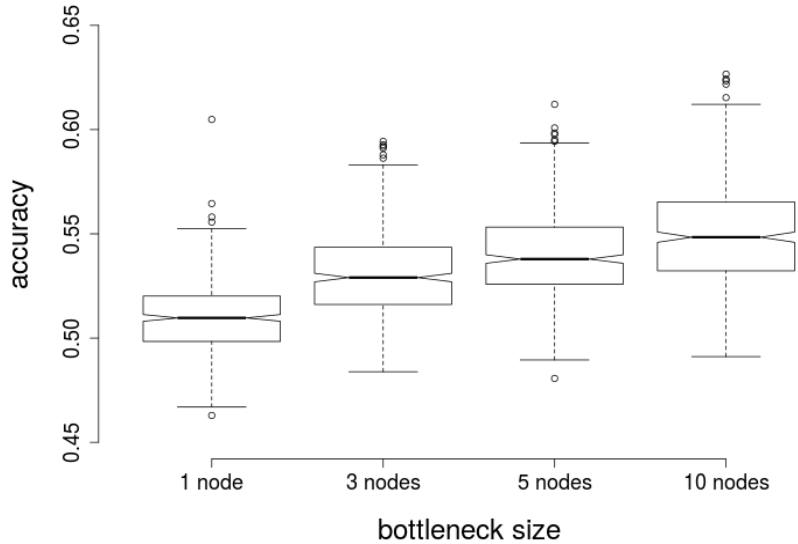}
\caption{Notched box-and-whiskers plots for accuracies for different bottleneck sizes within a feed-forward neural network model. A statistically significant difference in medians at 95\% confidence is indicated if notches are non-overlapping. The lower and upper ends of a box indicate the first and third quartile, while the median is depicted as a horizontal bar. Outliers are shown as circles above or below the whiskers.}
\label{fig:figure_5}
\end{center}
\end{figure}

In order to show the effect of bottleneck sizes, the box plots for four cases are depicted in Figure~\ref{fig:figure_5}. The models' accuracies significantly increase with the increases in bottleneck sizes, with the maximum (10 bottleneck nodes) resulting in an accuracy slightly below the full model without a bottleneck. The quartile ranges of the box plots, remaining approximately symmetrical, also increase with a higher number of bottleneck nodes. The inclusion of a bottleneck layer in the neural network model should hinder its performance, with the number of nodes forming the bottleneck being the deciding factor. However, if judged via the notches of the box plots, the step-wise increases from one to three, then five, and finally ten nodes, each time lead to a statistically significant rise in performance. While a one-node bottleneck results in an average accuracy of 51.07\%, the result for a ten-node bottleneck, with 55.03\%, differs only by 0.99\% from the accuracy of the same model and dataset without a bottleneck layer.

The possibility of the model not learning anything new after the introduction of bottleneck(s), meaning  the bottlenecked model performance being identical to a model with less hidden layers, has to be taken into account. This can, however, be safely dismissed due to the accuracies for 3, 5 and 10 nodes being notably different from each other. The results suggest that a large portion of the information can be compressed in ten weighted variables halfway through the model, which gives a rough indication of the overall complexity of the prediction problem itself. Notably,  Park et al. \cite{Park2007} find that stock market complexity has decreased over the course of the last 20 years, so future research may find smaller bottlenecks to be similarly viable.

\begin{figure}[!htb]
\begin{center}
% NOTE: FIGURES ARE OPTIMISED TO NOT SPAN THE FULL PAGE WIDTH (1.5-COLUMN FIT)
% NOTE: FIGURE SHOULD NOT BE PRINTED IN COLOUR
\includegraphics[width=0.77\columnwidth]{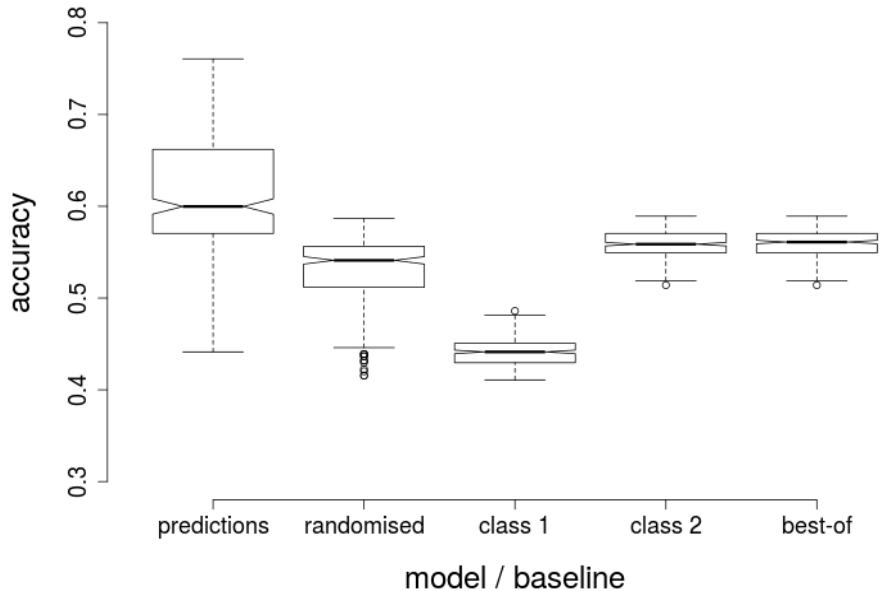}
\caption{Notched box-and-whiskers plots for accuracies in a high-volatility scenario. A statistically significant difference in medians at 95\% confidence is indicated if notches are non-overlapping. The lower and upper ends of a box indicate the first and third quartile, while the median is depicted as a horizontal bar. Outliers are shown as circles above or below the whiskers.}
\label{fig:figure_6}
\end{center}
\end{figure}

Figure~\ref{fig:figure_6} presents plots for accuracies computed in a high-volatility (crisis) environment. Table~\ref{tab:table_3} also shows the key performance estimates for the same scenario. Specifically, the results are for the trading environment during the global financial crisis of 2007/2008. The results show a large spread of the accuracies for different stocks. Notably, the medians and mean accuracies for one-class predictions show that the price trends more often change downwards during this modelled volatile scenario. In addition, the distribution of the model's accuracies is also skewed below the median; for example, the accuracies are spread higher upwards from the median, and the interquartile ranges are wider than for non-crisis scenarios.

\begin{table}[!htb]
\centering
\begin{tabular}{l l l l l}
\hline
\multicolumn{5}{l}{\textbf{accuracies of predictions}} \\
\hline\noalign{\smallskip}
model & rand. & class 1 & class 2 & best-of \\
\noalign{\smallskip}\hline\noalign{\smallskip}
$0.6113$ & $0.5301$ & $0.4405$ & $0.5595$ & $0.5595$ \\
\noalign{\smallskip}\hline\noalign{\smallskip}
\multicolumn{5}{l}{\textbf{tests against baselines}} \\
\noalign{\smallskip}\hline\noalign{\smallskip}
 & rand. & class 1 & class 2 & best-of \\
\noalign{\smallskip}\hline\noalign{\smallskip}
\textit{p}-value & $< 0.001$ & $< 0.001$ & $< 0.001$ & $< 0.001$ \\
min. diff. & $0.0677$ & $0.1589$ & $0.0399$ & $0.0388$ \\
\noalign{\smallskip}\hline
\end{tabular}
\caption{Statistical KPIs for high-volatility environments. In the table, accuracies, $p$-values and the minimal difference with regard to the significant difference in means are given. `class 1' is the prediction that stock trends will change downwards, and `class 2' is the prediction that stock trends will change upwards, whereas `randomised' represents the same prediction distribution as the model's predictions sampled randomly to test for simple distribution learning. The `best-of' features the respective highest accuracy of the three baselines, meaning `class 1', `class 2' and `randomised', to result in a fourth baseline that is, by default, at least 50\%.}
\label{tab:table_3}
\end{table}

While the average accuracies for predicting exclusively down- or upwards trend changes do not differ by more than 0.9\% for the primary experiments, this difference grows to 11.90\% for the crisis data. The model's high accuracy of 61.13\% can be partly explained by this difference, as the mean accuracy for predicting exclusively negative gradient changes is 55.95\% for the model's predictions, as shown in Table~\ref{tab:table_3}. The latter results are consistent with the general downwards-oriented trend of the market during a financial crisis, when confidence in the economy is expected to have fallen. Nevertheless, the additional accuracy of the model, along with the accuracy of the randomised mock predictions being below the exclusive predictions for negative trend changes, demonstrates that the model is able to exploit correlations in high-volatility environments such as the 2007/2008 financial crisis period. Thus, our findings are robust in the presence of extreme market stress.

\section{Conclusion}
\label{sec:conclusion}

In this paper, we add further evidence to the hypothesis that stock price data contains lagged correlations that can be successfully exploited through the application of deep learning models, including the exposure to financial crisis environments. Specifically, by using trend approximations as features, we can make higher-than-average price trend predictions without using the data of the target stock in the inputs. This finding contradicts the essence of the random walk hypothesis and the semi-strong and strong forms of the efficient market hypothesis, while underscoring the viability of applying deep learning for trend prediction in cross-correlated financial time series. Several tests of robustness are implemented, including testing the model in an environment of extreme price volatility, all of which firmly support the primary findings that stock price data contains actionable information through lagged correlations. In addition, the levels of prediction accuracies we obtain for the S\&P 500 stocks in our sample outperform the predefined baselines for strict statistical key performance indices to ensure significance.

Furthermore, we find that predictions of one stock's trend changes based on other stocks' price trend gradients in the preceding time interval improves in accuracy for larger time intervals, with average and maximum accuracies of 56.02\% and 63.95\%, respectively, for one-day predictions, all in excess of stated baseline predictions. The estimates retain large parts of their accuracy for a minimum of 10 nodes for mid-model bottleneck layers, and show equally above-baseline predictions in high-volatility market scenarios, albeit with the trade-off of a higher variance for different stocks. The large difference in accuracies for the model's predictions and the baselines cannot be explained by the shift to a skewed distribution in favour of negative trend changes, given that the training data does not contain such an imbalance. An alternative explanation that assumes market shocks steering large-scale trend movements is ruled out by testing the success of randomised model predictions.

As discussed previously by Moews et al. \cite{Moews2019}, weak-form market efficiency can be upheld under the additional assumption that only a select number of market agents have access to, or choose to implement, techniques such as the one described in this work. In other words, if a small-enough amount of capital is involved with exploiting lagged correlations in stock markets, these correlations are still present in price data and the market remains efficient for the majority of market agents. Given the observed reluctance of financial firms to deploy artificial intelligence approaches on a large scale, as well as the inherent difficulties of black-box models when having to produce research for market oversight agencies, this scenario does not seem unlikely and allows for our results to be compliant with much of the financial literature.

This interpretation resembles the adaptive market hypothesis put forward by Lo \cite{Lo2004}, which extends the efficient market hypothesis through behavioural economics. One of the central tenets of this extension is that reduced market efficiency results from a lack in competition for abundant resources, whereas scarce resources in highly competitive environments result in more effective markets. For the U.S. market, this hypothesis is, for example, supported by research by Ito and Sugiyama \cite{Ito2009} and Kim et al. \cite{Kim2011}. Further, more compelling evidence for the adaptive market hypothesis is subsequently presented by Urquhart and McGroarty \cite{Urquhart2014, Urquhart2016}, using long-term datasets for the Dow Jones Industrial Average and various global stock exchanges, respectively, and through investigations of bond markets \cite{Charfeddine2018}.

Finally, the results in this paper demonstrate the value of deep learning approaches to both risk diversification and econometrics, and show the utility of linear regression derivatives as features for time series-based investigations, thus offering a simple trend indicator with a high predictive value in order to further the understanding of large-scale time series correlations in complex systems.

\section*{Acknowledgements}

We wish to express our gratitude to Peter Zimmerman and other discussants of this paper at the Third Workshop of the European Capital Markets Cooperative Research Centre (ECMCRC) on July 5 2019 in Dublin, Ireland, whose additional insights led to the improvement of this work.

%\section*{Funding}
% No funding was provided for, or is associated with, this work.

% Bibliography
\section*{References}

\bibliographystyle{elsarticle-num}
\bibliography{References.bib}

\end{document}